      [1999/12/01 v1.4c Il Nuovo Cimento]
\documentclass{cimento2}

\usepackage{graphicx}  % got figures? uncomment this
\title{Two and a half years of GRB localizations with the INTEGRAL
Burst Alert System}
\author{S.~Mereghetti
and
 D.~G\"{o}tz
}
 \instlist{\inst{} INAF - Istituto di Astrofisica Spaziale e
Fisica Cosmica, sezione di Milano - via E.Bassini 15, 20133
Milano, Italy
                          }
\PACSes{{ }{\textit{A shorter version of this paper will be
published in the Proceedings of the 4$^{th}$ Workshop "Gamma-Ray
Bursts in the Afterglow Era", Roma, 2004 October 18-22, eds. L.
Piro, L. Amati, S. Covino, and B. Gendre. Il Nuovo Cimento, in
press}}
}
\begin{document}

%  \special{!userdict begin /bop-hook{gsave 150 90 translate
%  55 rotate /Times-Roman findfont 60 scalefont setfont
%  0 0 moveto 0.7 setgray (Draft CONFIDENTIAL - 27/5/05) show grestore}def end}

\maketitle

\begin{abstract}
We review the results on Gamma-ray Bursts obtained during the
first two and a half years of operations of the INTEGRAL Burst
Alert System (IBAS).  In many cases GRB coordinates have been
distributed with an unprecedented combination   of accuracy (3$'$)
and speed (20-30 s).  The resulting rapid follow-ups at other
wavelengths, including sensitive \textit{XMM-Newton} and
\textit{Swift} observations, have led to several interesting
results.
\end{abstract}

\section{Introduction}

The INTEGRAL satellite, launched on October 17, 2002, has been
designed as a general purpose mission dedicated  to
high-resolution imaging and spectroscopy in the hard X--ray / soft
$\gamma$-ray energy range.  Thanks to the good imaging
capabilities of its IBIS instrument \cite{ibis} and the continuous
data transmission to the ground, it has been possible to set up an
automatic system for the rapid distribution of GRB coordinates. In
the last two and a half years, the INTEGRAL Burst Alert Systems
(IBAS, \cite{ibas}) has provided some of the fastest and most
accurate localizations ever obtained for GRBs. In some cases the
GRB coordinates were distributed while the gamma-ray emission was
still ongoing. Whenever a rapid follow up at X--ray wavelengths
has been carried out, mostly with \textit{XMM-Newton}, an
afterglow has been detected, often leading to very interesting
results. After a brief description of the IBAS system, we
summarize the global properties of the 28 GRBs localized to date
(May 2005) and discuss a few of the most interesting cases.

\section{The INTEGRAL Burst Alert System}

Thanks to its 72 hours-long and highly eccentric  orbit, the
INTEGRAL satellite is in continuous contact with the ground
stations during the observations. Therefore the data are
transmitted to ground without significant delays. This has allowed
us to implement a software system for the automatic GRB search at
the INTEGRAL Science Data Center (ISDC \cite{isdc}), where the
data are received after only a few seconds. The IBAS rapid
localizations are based on data of the imager instrument IBIS, and
in particular of its ISGRI detector \cite{isgri}, which offers the
best performances in terms of large field of view
(29$^{\circ}\times$29$^{\circ}$) and angular resolution over the
20 keV - 1 MeV energy range, well suited for the detection of
GRBs. The search for GRBs is done in parallel by several programs
using different energy ranges (between 15 and 200 keV), triggering
time scales (between 8 ms and 100 s), and imaging methods (for
more details see \cite{ibas} and
\textit{http://ibas.mi.iasf.cnr.it}).

Typically the IBAS localizations obtained by the automatic
software and distributed in real time have 90\% confidence level
error radii of $\sim3'$. These errors can be reduced down to
$\sim2'$ or less, depending on the source signal to noise ratio,
in the subsequent off-line analysis. The GRB positions derived by
IBAS are delivered via Internet to all the interested users. For
the GRBs detected with high significance, this is done immediately
by the software which sends \textit{Alert Packets} using the UDP
transport protocol. In case of events with lower statistical
significance, the alerts are sent only to the members of the IBAS
Localization Team, who perform further analysis and, if the GRB is
confirmed, can distribute its position with an \textit{Off-line
Alert Packet}.

%--------------------------------------------------------
\begin{table}[ht!]
\begin{center}
\caption[]{Properties of GRBs detected with IBAS}
\begin{tabular}{lcccc}
\hline \noalign {\smallskip}
GRB      &  Duration  & Peak Flux        & Power law                & Afterglows$^{b}$ \\
         & T$_{90}$  & (20-200 keV)     & photon index$^{a}$     &   \\
       &  [s]  & [photons cm$^{-2}$s$^{-1}$] &                     &   \\
\hline \noalign {\smallskip}
GRB 021125   & 25  &   22  & 2.2 \cite{021125P}            &    --  \\
GRB 021219   & 5.5 &  4.0  & 2.0 \cite{021219P}  & --  \\
GRB 030131   & 124 &  1.9  & $\sim2$ \cite{030131P}    & O \cite{030131O} \\
GRB 030227   &  33 &   1.1 & 1.9 \cite{030227P}   &  O \cite{030227O}, X \cite{030227P,watson} \\
GRB 030320   &  48 &   3.6 & 1.7 \cite{030320P} & --   \\
GRB 030501   &  40 &   2.7 & 1.75 \cite{030501P}  & -- \\
GRB 030529   &  20 &   0.4 & 1.7 $^{c}$      &  -- \\
GRB 031203   &  39 &   1.7 & 1.63 \cite{031203P} &   R \cite{031203R}, O \cite{031203O, malesani}, X \cite{031203X,ring}  \\
GRB 040106   &  47 &   1.0 & 1.7 \cite{040106P} & R \cite{040106O}, X \cite{040106X,040106P} \\
GRB 040223   & 258 &   0.4 & 2.3 \cite{040223}&   X  \cite{040223X} \\
GRB 040323   &  14 &   1.6 &    $^{d}$        &  O\cite{040323O}  \\
GRB 040403   &  21 &   0.5 & 1.9 \cite{040403P} & --          \\
GRB 040422   &  10 &   3.5 &    $^{d}$        & --  \\
GRB 040624   &  35 &   0.5 &    $^{d}$        & --    \\
GRB 040730   &  43 &   0.4 &    $^{d}$        &  --   \\
XRF 040812   &  19 &   0.6 &    2.4 \cite{040812}        & O\?, X \cite{040812X}   \\
GRB 040827   &  49 &   0.6 &    $^{d}$        & NIR \cite{040827X}, X \cite{040827X}   \\
XRF 040903   &  26 &   0.4 &   2.9 \cite{040903P}        & --  \\
GRB 041015   &  30 &   0.2 &    $^{d}$        & --    \\
GRB 041218   &  40 &   3.0 &    $^{d}$        &  O \cite{041218O}  \\
GRB 041219A   & 560 & $>$12 &    $^{d}$        &   O \cite{041219O}, NIR \cite{041219IR}   \\
GRB 050129   &  80 &  0.26 &    $^{d}$        &   --  \\
GRB 050223   &  51 &   0.6 &    $^{d}$        & X \cite{050223X}      \\
GRB 050502A  &  16 &   1.8 &    $^{d}$        & O \cite{050502O}, NIR \cite{050502I}    \\
GRB 050504   &  108 &   0.5 &    $^{d}$        &   X \cite{050504X} \\
GRB 050520   &  58 &   0.7 &    $^{d}$        &    X \cite{050520X} \\
XRF 050522   &  11 &   0.34&    $^{d}$         &  X \cite{050522X}  \\
GRB 050525A  &   9 &   33.4&    2.1$^{c}$         & R \cite{050525R}, O \cite{050525Oa,050525Ob}, X \cite {050525D} \\
% \noalign {\smallskip} \hline \label{tab:spec}
\hline
\end{tabular}
\end{center}
 $^{a}$  In the 20-200 keV range a power law is adequate to fit most INTEGRAL GRBs
  % The two values for GRB 021125 are for the ranges 20-200 keV (ISGRI) and 170-500 keV (PICsIt). The arrows
% indicate time evolution.

$^{b}$ O=optical, R=radio, X=X-rays

$^{c}$ This work

$^{d}$ Results not yet published

\end{table}
%----------------------------------------------------------------

\begin{figure}
\begin{center}
\includegraphics[width=8cm]{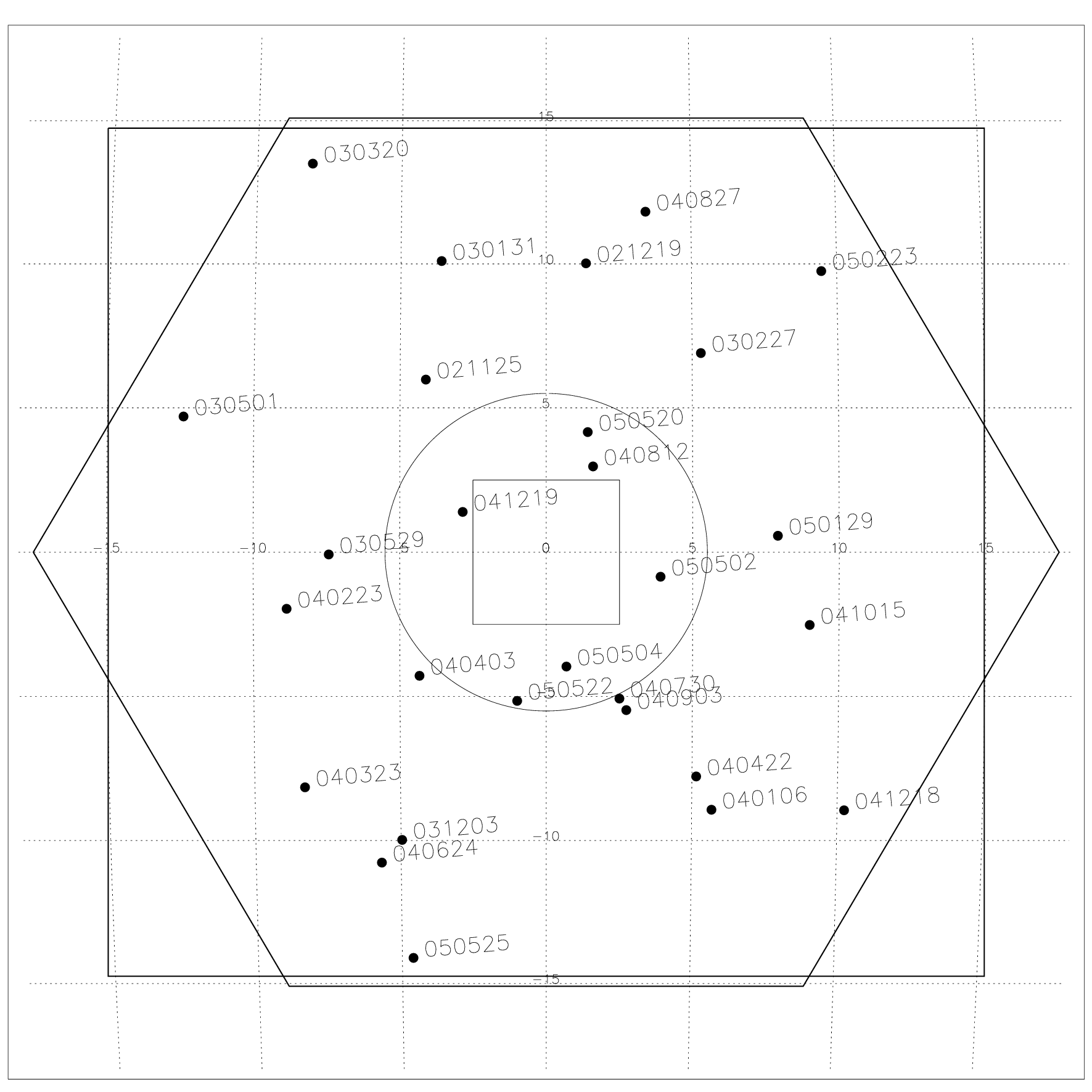}
\caption{Positions of the GRBs localized with IBAS in the fields
of view of the INTEGRAL instruments:  IBIS (large square), SPI
(hexagon), JEM-X (circle), OMC (small square).}
\end{center}
\end{figure}

\begin{figure}
\begin{center}
\includegraphics[width=12cm]{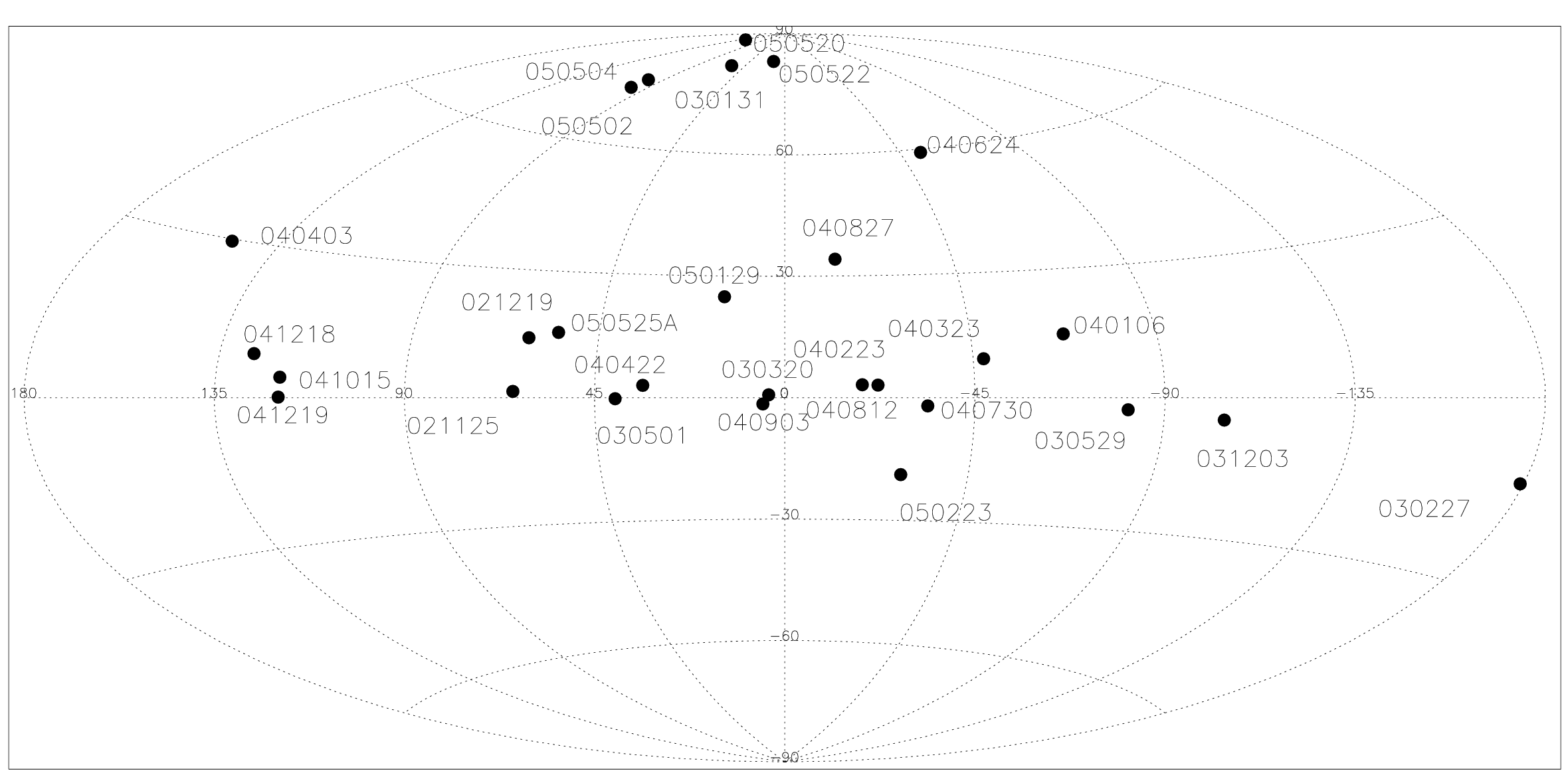}
\caption{Positions in Galactic coordinates of the GRBs localized
with IBAS. The spatial distribution reflects the highly
non-uniform sky coverage of the INTEGRAL observations.}
\end{center}
\end{figure}

\section{The INTEGRAL GRB sample}

Up to now (May 2005), 28 GRBs have been discovered in the field of
view of IBIS. Figure 1 shows their positions in the fields of view
of the INTEGRAL instruments. Six of them were within the field of
view of the X-ray monitor JEM-X \cite{jemx} and thus detected also
below 20 keV. Unfortunately none of them was sufficiently on axis
to be also observed with the optical camera OMC \cite{omc}.

The main properties of the INTEGRAL GRBs are summarized in
Table~1. The peak fluxes in the 20-200 keV (for an integration
time of 1 s) have been recomputed using the latest available
calibrations and supersede previously reported values. They are
typically in the range $\sim$0.3--5 photons cm$^{-2}$ s$^{-1}$,
with a few exceptions like GRB 021125, the first GRB detected by
INTEGRAL \cite{021125P}, GRB 041219, for which only a lower limit
on the peak flux could be derived since it was so bright to
saturate the available IBIS telemetry \cite{041219}, and GRB
050525A, which was discovered by \textit{Swift} \cite{050525D}.
Spectral information for many of the most recently detected bursts
has not been published yet. However, the results available so far
indicate a relatively  large proportion of bursts with soft
spectra: 9 of the 14 bursts with published spectral information
(Table 1)  have a 20-200 keV spectrum steeper than 1.75, which
qualifies them as X-ray rich GRBs and/or X-ray flashes.  The
latter category comprises at least XRF 040812 \cite{040812}, XRF
040903 \cite{040903P}, and XRF 050522 \cite{050522D}.

X-ray follow-ups have been carried out with \textit{XMM-Newton}
for six GRBs   and with \textit{Chandra} for GRB 040812, typically
starting about 5-8 hours after the GRB. In all these cases an
X--ray afterglow has been found within the IBAS error regions.
More rapid X-ray follow-ups with \textit{Swift}  have been done
for all the GRBs detected after April 2005, leading in all but one
case (GRB 050502A \cite{050502X}) to the discovery of an
afterglow.

Several bursts occurred in heavily absorbed regions of the
Galactic plane, which is the sky region most observed by INTEGRAL
(see Fig.~2). Optical and/or IR afterglows  have been found for 10
GRBs, and a few other candidate counterparts have been reported.
The particularly interesting case of the prompt optical emission
observed simultaneously with the gamma-rays in  GRB 041219A is
discussed in Section 4.7.  The host galaxy has been identified for
GRB 031203 \cite{031203O} at a remarkably small redshift of z=0.1,
while a redshift z=3.79 has been tentatively derived for GRB
050502A \cite{050502Z}. In a few cases the optical observations
were deep enough to provide interesting upper limits at early
times. Striking examples are GRB 050520 (Sect. 4.11) and GRB
040403 \cite{040403P}.

The overall picture indicates that the INTEGRAL sample  includes
many  faint and X-ray rich bursts which are also relatively dim in
the optical. Indeed the bursts localized by IBAS are among the
faintest ones for which good localizations have been obtained
before the \textit{Swift} era. This is shown in Fig.~3, where we
compare the peak fluxes of the GRBs localized by  INTEGRAL and
\textit{BeppoSAX}.

\begin{figure}
\begin{center}
\includegraphics[width=12cm]{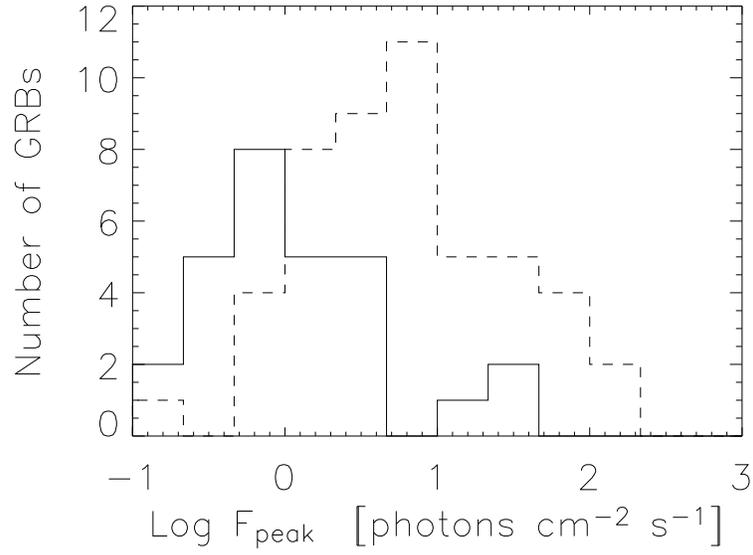}     % includes figure foo.eps
\caption{Distribution of the peak fluxes for the GRBs localized
with IBAS (solid line) and with \textit{BeppoSAX} (\cite{sax},
dashed line).}
\end{center}
\end{figure}

\begin{figure}
\begin{center}
\includegraphics[width=14cm]{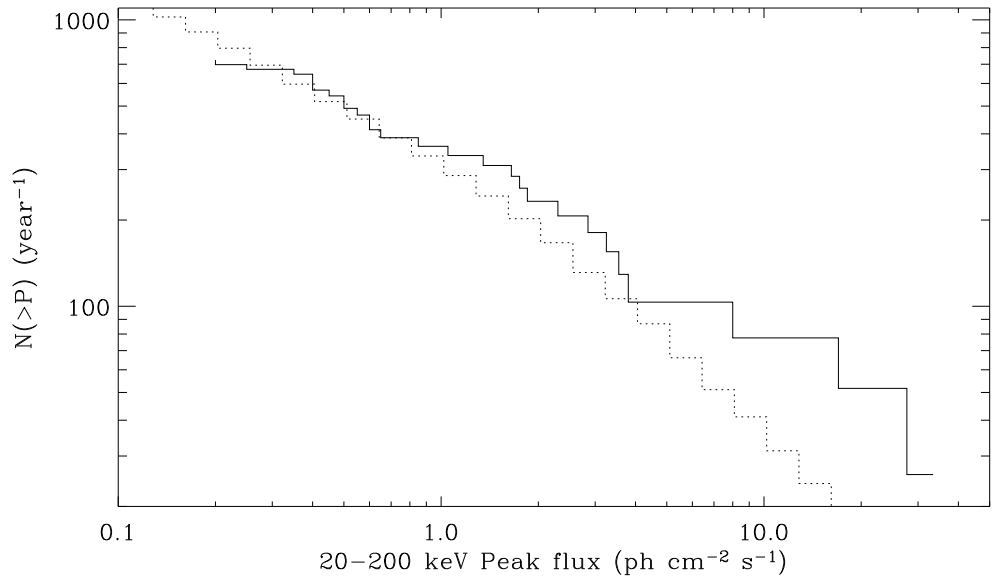}     % includes figure foo.eps
\caption{LogN-LogP distribution of the INTEGRAL bursts. The dashed
line is the BATSE logN-logP \cite{sah} converted to the 20-200 keV
energy range.}
\end{center}
\end{figure}

Despite the statistics is still limited, it is possible to derive
a first estimate of the distribution of peak fluxes (LogN-LogP
curve) of the 28 GRBs detected to date in the IBIS field of view
($\sim$0.27 sr, i.e. 2\% of the sky). Considering the time spent
below the radiation belts, the satellite slews (during which IBAS
is not active), periods of high solar activity, and gaps due to
operational and/or technical problems, we estimate that the
effective exposure reached with IBAS up to now corresponds to
$\sim$1.8 years. This value has been used to normalize the
LogN-LogP shown in Fig.~4 (note that the error on the
normalization is dominated by the small statistics of the sample,
therefore a more accurate estimate of the effective exposure is
not required at this stage). The INTEGRAL  LogN-LogP refers to the
peak fluxes in the 20-200 keV energy range. To compare it with the
BATSE results obtained in the 50-300 keV range, we converted the
BATSE LogN-LogP \cite{sah} to our energy interval assuming an
average spectral shape given by a Band function with $\alpha$=--1,
$\beta$=--2.5, and E$_{peak}$=200 keV (dashed line). The BATSE and
INTEGRAL LogN-LogP distributions are in good agreement.

\section{Results on individual GRBs}

\subsection{GRB 030227}

This is the first INTEGRAL GRB for which  X--rays and optical
afterglow searches were successful. An \textit{XMM-Newton} Target
of Opportunity Observation could start  only  8 hours after the
GRB, leading to the discovery of an  X--ray afterglow with  0.2-10
keV flux decreasing as t$^{-1}$ from 1.3$\times$10$^{-12}$ to
5$\times$10$^{-13}$ erg cm$^{-2}$ s$^{-1}$. The afterglow spectrum
was well described by a power law with photon index 1.94$\pm$0.05
and absorption of a few times 10$^{22}$ cm$^{-2}$, significantly
larger than the Galactic value \cite{030227P}.  This supports the
scenarios involving the occurrence of GRBs in regions of star
formation. Some evidence for an emission line at 1.67 keV, which
if attributed to Fe would imply a redshift $z\sim$3, was also
found in the \textit{XMM-Newton} spectrum \cite{030227P}. However,
this is inconsistent with another claim for lines possibly
appearing toward the end of the same \textit{XMM-Newton}
observation \cite{watson},  interpreted as H- and He-like lines
from Mg, Si, S, Ar and Ca at z=1.39. The   statistical
significance of these spectral features is debated, as in the case
of similar claims in other GRBs.
% (see e.g. \cite{sako}).
The much more rapid X--ray follow ups which can now be obtained
with \textit{Swift} \cite{swift} will certainly settle this
important issue in the coming months.

\subsection{GRB 030529}

This burst has been detected during an off-line reprocessing of
the first months of INTEGRAL data, which was performed in order to
scan also the older data with the most recent and sensitive
versions of the IBAS programs. The properties of GRB 030529, the
only burst found in this off-line IBAS reprocessing, are reported
here for the first time. It occurred at 19:53:15 UT at coordinates
R.A.= 09$^h$ 40$^m$ 29.3$^s$, Dec.=--56$^{\circ}$ 20$'$ 31$''$
(J2000,  uncertainty  3$'$). It was a faint burst, lasting about
20 s, with a power law spectrum with photon index
$\Gamma$=1.71$\pm$0.20, and fluence  4$\times$10$^{-7}$ erg
cm$^{-2}$ (20-200 keV). The instrumental background at the time of
the burst detection was highly variable. This explains why GRB
030529 was missed by the earlier versions of the IBAS programs.

\subsection{GRB 031203}

Also for this GRB very interesting results could be obtained
thanks to the   X-ray, radio  and optical follow-up observations
enabled by the rapid IBAS localization. This is one of the few
GRBs for which there is spectroscopical evidence of an associated
Type Ic  Supernova \cite{malesani}. The discovery of its host
galaxy led to a redshift determination of z=0.1 \cite{031203O},
making GRB 031203 the second closest GRB, and implying a
surprisingly small isotropic-equivalent  energy
E$_{iso}$=(6--14)$\times10^{49}$ erg s$^{-1}$ \cite{031203P}. This
value and the lower limit on E$_{peak}$ ($\sim$200 keV), derived
from the average spectrum, make GRB 031203 an outlier in the
E$_{peak}$-E$_{iso}$ relation \cite{Amati}.

The X--ray images obtained with \textit{XMM-Newton} led to the
discovery of an expanding ring due to the scattering of the GRB
X-ray emission by dust grains in our Galaxy \cite{ring}. The
modelling of this GRB ``echo'' gives us an indirect mean to
estimate the intensity of the prompt GRB  emission at  X--ray
energies. This gives some evidence for an X-ray flux component in
excess of the low-energy extrapolation of the INTEGRAL spectrum
\cite{thisosb}.

\subsection{GRB 040106}

The afterglow of this GRB has been promptly (only 5 hours later)
and deeply observed in the X-rays by \textit{XMM-Newton}: its
1--10 keV spectrum is uniquely hard (power law photon index
1.49$\pm$0.03), and  its  temporal decay is a power law of index
1.46$\pm$0.04  \cite{040106P}. Assuming that the cooling frequency
$\nu_{c}$ is below the observed X--ray range (as suggested by the
optical data), these values do not fit with any  of the simple
fireball models. On the other hand, if the $\nu_{X}<\nu_{c}$
regime applies, the afterglow can be described as a spherical
fireball expanding into a wind environment \cite{040106X}.

\subsection{GRB 040403}

GRB 040403 is one of the faintest gamma-ray bursts for which a
rapid (30 s) and accurate (2.8$'$) localization has been obtained.
Its  steep spectrum obtained with IBIS/ISGRI in the 20-200 keV
range (power law photon index = 1.9) implies that GRB 040403 is
most likely an X-ray rich burst \cite{040403P}.

Despite being at a Galactic latitude (b=30$^{\circ}$) higher than
the majority of INTEGRAL bursts and thus barely affected by
interstellar extinction (A$_{V}\sim$0.3), optical follow-ups were
somewhat discouraged by the presence of the full Moon.
Nevertheless a relatively deep limit of R$>$24.2 at 16.5 hours
after the burst could be obtained \cite{040403P}, indicating a
rather faint afterglow, similar to those seen in other soft and
faint bursts (e.g. GRB 030227 \cite{030227P}).

\begin{figure}
\begin{center}
\includegraphics[width=14cm,angle=0]{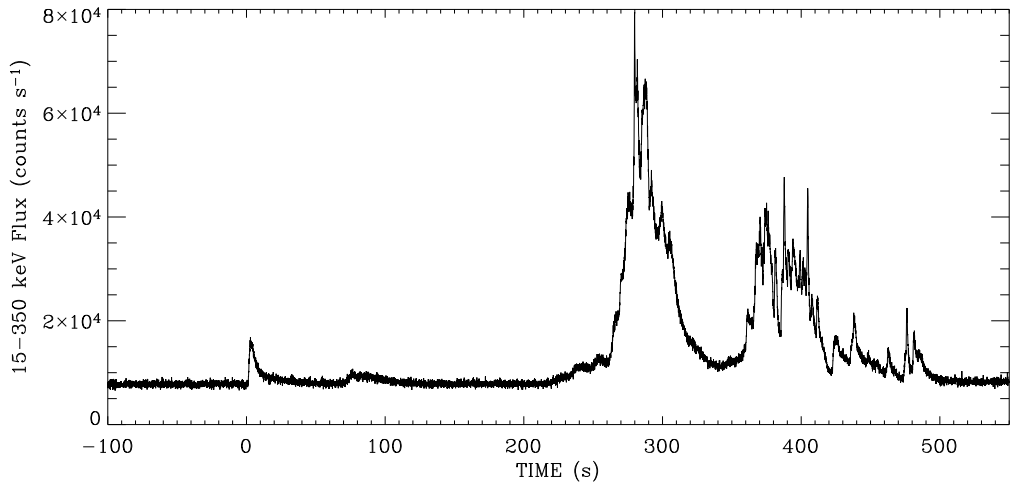}     % includes figure foo.eps
\caption{Swift light curve of GRB 041219A obtained with the BAT
instrument in the 15-300 keV energy range (data from
\textit{http://swift.gsfc.nasa.gov/docs/swift/swiftsc.html}). }
\end{center}
\end{figure}

\subsection{XRF 040903}

This faint burst had a very soft spectrum (it was not
significantly detected by IBIS above 60 keV), it came from a
direction close to the Galactic center (l=5.2$^{\circ}$,
b=--1.5$^{\circ}$), and a faint \textit{ROSAT} X-ray source was
present in its error region. For these reasons also the
possibility that it was due to a Type I X-ray burst from an
unidentified low mass X-ray binary was  considered in the initial
reports \cite{040903D}. However, subsequent analysis indicate that
a Type I X-ray burst origin is less likely \cite{040903P}. In
fact, its 20-100 keV spectrum is well fit by a power law with
photon index $\Gamma$=2.9$\pm$0.4, while the temperature obtained
with a blackbody fit (kT=6.9$\pm$1.5 keV) is much higher than
those typically seen in X-ray bursts. Contrary to the X-ray bursts
no spectral softening is visible in XRF 040903. Finally, no
persistent emission has been seen with INTEGRAL from the
\textit{ROSAT} source before or after the burst.

\subsection{GRB 041219A}

This bright and long GRB was discovered by IBAS and also observed
by \textit{Swift}, during its commissioning phase while it was not
yet distributing GRB localizations in real time. The light curve
obtained with the \textit{Swift} BAT instrument is shown in
Fig.~5. This burst started with a fainter precursor (at t=0 in
Fig.~5), followed by a nearly quiescent time interval of $\sim$200
s, before the main emission. The precursor peak was bright enough
to trigger IBAS and provide an accurate localization, which was
promptly distributed allowing the detection of optical
\cite{041219O} and NIR \cite{041219IR} flashes during the
gamma-ray prompt emission phase. A similar simultaneous detection
had been obtained before only for GRB 990123 \cite{990123}, which,
contrary to GRB 041219A, showed an anti-correlation between the
optical and gamma-ray light curves. This was interpreted as
evidence for an external reverse shock origin of the optical
emission. In GRB 041219A the optical flux intensity was instead
correlated with that at high-energy, suggesting that it was due to
internal shocks \cite{041219O}.

\subsection{GRB 050223}

This GRB was discovered by \textit{Swift} \cite{050223D}. Just by
chance INTEGRAL was observing in the direction of this burst, that
was thus detected by the IBIS instrument at an off-axis angle of
14$^{\circ}$ \cite{050223}. The burst triggered the IBAS programs,
however the corresponding \textit{Alert Packet} was not
distributed by the software, because the burst occurred during a
short  time interval (200 s) in which  one of the 8 modules
composing the IBIS/ISGRI detector was switched off (this is done
autonomously by the on board software handling the noisy pixels).
Under these circumstances there is the possibility that IBAS
incorrectly assigns to the GRB the coordinates of one of its ghost
images, therefore the \textit{Alert Packets} are not sent. In the
off-line analysis it was possible to take these effects into
account and results consistent with the \textit{Swift} ones were
obtained \cite{050223}.

\begin{figure}
\begin{center}
\includegraphics[width=12cm,angle=0]{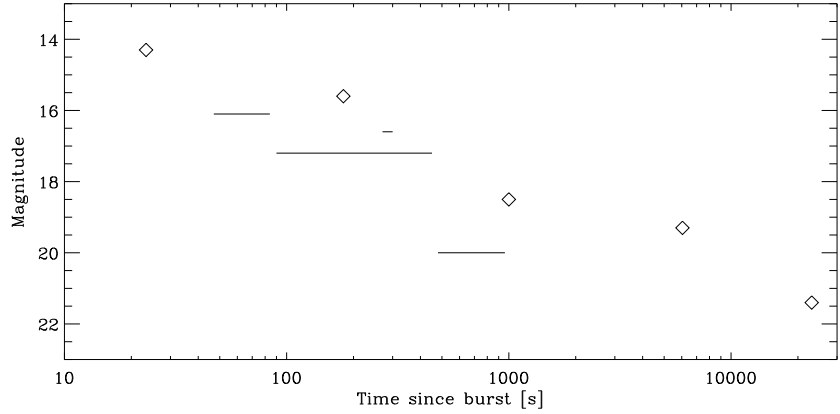}     % includes figure foo.eps
 \caption{\textit{Squares:} light curve of the early phase of the optical
 afterglow of GRB 050502A \cite{050502O,050502Ob,050502Oc}.
\textit{Lines:} Upper limits on the optical afterglow emission
from GRB 050520 \cite{050520Oa,050520Ob}}
\end{center}
 \end{figure}

\subsection{GRB 050502A}

The rapid IBAS localization of GRB 050502A allowed the detection
of the early part of the optical afterglow  \cite{050502O}, which
had magnitude 14.3 at 23 s after the burst and then faded quickly,
reaching magnitude 18 after 1000 s (Fig.~6). The afterglow was
faint also at X-ray energies: only an upper limit of
$\sim7\times10^{-15}$ erg cm$^{-2}$ s$^{-1}$ could be set with
\textit{Swift} \cite{050502X}, contrary to all the other INTEGRAL
bursts that were detected in X-rays by \textit{Swift}.
Observations with the Keck telescope were performed three hours
after the burst, leading to the determination of a redshift
z=3.793 \cite{050502Z}. If confirmed, this would be the third
highest redshift measured for a GRB.

\begin{figure}
\begin{center}
\includegraphics[width=8cm,angle=-90]{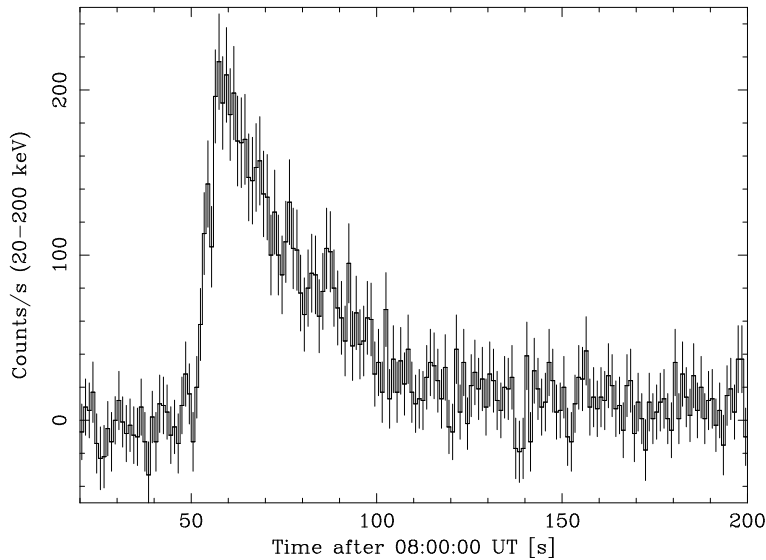}     % includes figure foo.eps
\caption{IBIS/ISGRI light curve of GRB 050504 in the 20-200 keV
energy range (from \textit{http://ibas.mi.iasf.cnr.it}). }
\end{center}
\end{figure}

\subsection{GRB 050504}

The light curve of GRB 050504, showing a typical fast-rise
exponential-decay,  is displayed  in Fig.~7. Despite the location
of this burst at high Galactic latitude (b=$+75^{\circ}$), the
presence of a 10$^{th}$ magnitude K type  star prevented the
observation of part of the IBAS error region (90$''$ radius).
However, the precise position derived for the X-ray afterglow
later discovered with \textit{Swift} \cite{050504X} is
inconsistent with that of the bright star. The upper limits for
the optical afterglow are thus rather constraining:  independent
groups reported values of R$\sim$19 \cite{050504Oa} and R$\sim$21
\cite{050504Ob} a few minutes after the burst.

\subsection{GRB 050520}

Also this multi-peaked GRB (Fig.~8) was localized by IBAS in real
time, leading to prompt observations with robot telescopes, which
however failed to see an afterglow. The upper limits
\cite{050520Oa,050520Ob} are compared to the positive detection of
the faint afterglow of GRB 050502A in Fig.~6, and indicate that
GRB 0505020 was really optically dark. Note that these two bursts
had similar values of the peak flux (Tab. 1) although different
durations and time profiles. A faint X-ray afterglow was detected
with \textit{Swift} \cite{050520X}.

\begin{figure}
\begin{center}
\includegraphics[width=8cm,angle=-90]{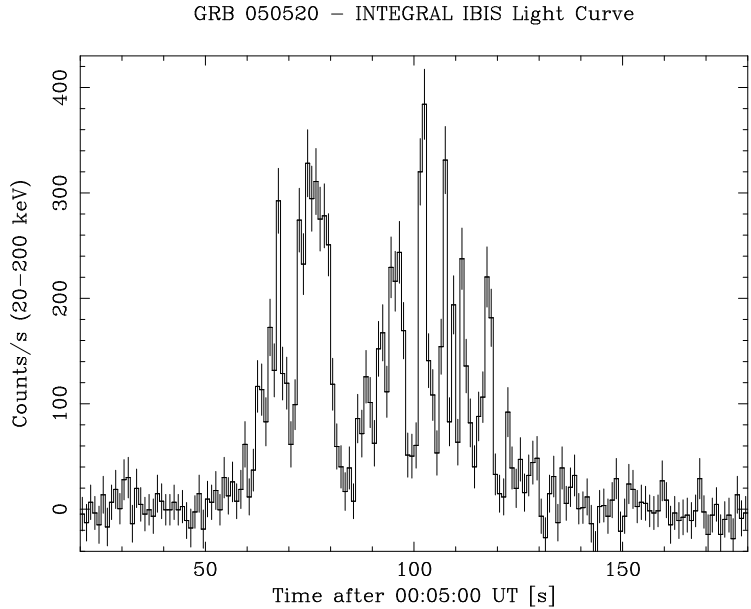}     % includes figure foo.eps
\caption{IBIS/ISGRI light curve of GRB 050520 in the 20-200 keV
energy range (from \textit{http://ibas.mi.iasf.cnr.it}). }
\end{center}
\end{figure}

\subsection{XRF 050522}

No emission above  ~40 keV was detected with IBIS/ISGRI for this
burst that can thus be classified as an X--ray flash
\cite{050522D}. \textit{Swift} reported an X--ray source outside
the IBAS error region, but subsequent observations showed it to be
unrelated to the burst and led to the discovery of a fading X--ray
afterglow inside the error box \cite{050522X}. This is one of the
faintest bursts detected with INTEGRAL. Its light curve is shown
in Fig.~9.

\begin{figure}
\begin{center}
\includegraphics[width=8cm,angle=-90]{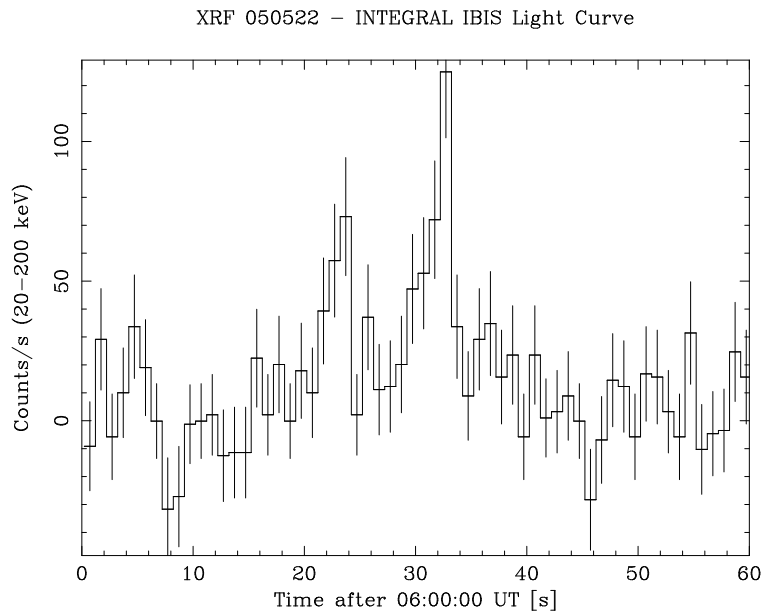}     % includes figure foo.eps
\caption{IBIS/ISGRI light curve of XRF 050522 in the 20-200 keV
energy range (from \textit{http://ibas.mi.iasf.cnr.it}). }
\end{center}
\end{figure}

\begin{figure}
\begin{center}
 \includegraphics[width=8cm,angle=-90]{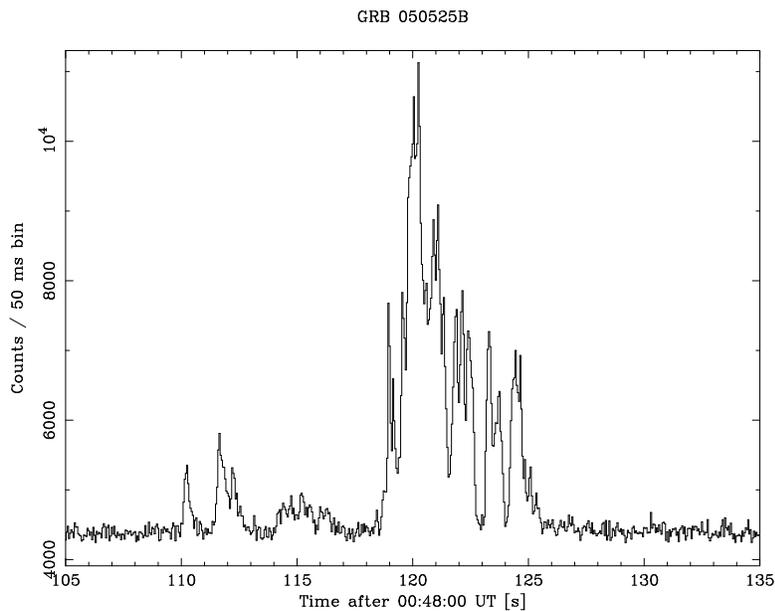}     % includes figure foo.eps
 \caption{Light curve of GRB 050525B obtained with the
 INTEGRAL SPI Anti-Coincidence Shield, which is sensitive to photons of energy
 above $\sim$80 keV.}
\end{center}
 \end{figure}

\subsection{GRB 050525A}

This very bright burst, discovered by \textit{Swift}
\cite{050525D}
%which distributed its position
%with a 6 minute delay \cite{050525D}. It was just
occurred by chance within the  IBIS field of view \cite{050525},
although at a very large off-axis angle  (see Fig.~1). The large
coding noise in the deconvolved image, due to the extremely
off-axis position (14.7$^{\circ}$), and the simultaneous presence
of Cyg X-1 in the field of view, conspired to make GRB 050525A
undetectable in real time by the IBAS programs. This was
unfortunate, since   the \textit{Swift} localization had a delay
of six minutes and the initial part of the optical afterglow was
missed by ground based telescopes. The  ROTSE-III and TAROT robot
telescopes detected the afterglow at R$\sim$15 ten minutes after
the burst \cite{050525Oa,050525Ob}, while earlier optical data
were obtained with UVOT on board \textit{Swift} \cite{050525U}.
Further observations led to the determination of the redshift
z=0.6 \cite{050525Z} and to the detection of the radio afterglow
\cite{050525R}.

GRB 050525A is the brightest burst detected to date in the field
of view of IBIS. A preliminary analysis of the IBIS/ISGRI data
indicates that its 20-200 keV spectrum can be fit by a single power law
with photon index 2.1$\pm$0.1 (note
that the instrument calibration at such a large off-axis angle is
still subject to some uncertainties).

\subsection{GRB 050525B}

GRB 050525A was also seen with the \textit{Konus-Wind} satellite
\cite{050525K}, which, quite surprisingly, detected an even
brighter burst 45 minutes later. Since the arrival direction of
the second burst could not be determined, considering the small
probability of seeing two very bright events so close in time, the
possibility that the two were related was suggested
\cite{050525K}. However, this was excluded by INTEGRAL data, since
at the time of the second burst, GRB 050525B \cite{050525B}, no
signal was detected from the sky position of GRB 050525A, that was
now well inside the IBIS field of view. GRB 050525B was clearly
seen in the Anti-Coincidence Shield \cite{acs} of the INTEGRAL SPI
instrument (see Fig.10).

\acknowledgments
 IBAS is successfully working thanks also  to the continuous support of J.Borkowski,
 M.Beck, N.Mowlavi, S.Shaw and all the operational staff of the ISDC and MOC.
This paper is based on observations with INTEGRAL, an ESA project with instruments
and science data centre funded by ESA member states (especially the PI countries:
Denmark, France, Germany, Italy, Switzerland, Spain), Czech Republic and Poland,
and with the participation of Russia and the USA.

\end{document}